\documentclass[a4paper,pra,twocolumn,superscriptaddress]{revtex4-2}

\usepackage{amssymb}
\usepackage{amsmath}
\usepackage{color}
\usepackage{graphicx}
\usepackage{bbold}
\usepackage{float}
\usepackage{braket}
\usepackage[colorlinks,citecolor=blue]{hyperref}
\usepackage{subfigure}
\begin{document}
	
	\date{\today}
	\title{Breakdown of Aharonov-Bohm cage in Rydberg synthetic lattices: the roles of inhomogeneity and long-range exchange}
	\author{Li Pan}
	\affiliation{Institute of Atomic and Molecular Physics, Sichuan University, Chengdu 610065, China}
	\affiliation{College of Physics, Sichuan University, Chengdu 610065, China}
	
	\author{Xinlu Chen}
	\affiliation{Institute of Atomic and Molecular Physics, Sichuan University, Chengdu 610065, China}
	\affiliation{College of Physics, Sichuan University, Chengdu 610065, China}
	
	\author{Hong Zhang}
	\affiliation{College of Physics, Sichuan University, Chengdu 610065, China}
	\affiliation{Key Laboratory of High Energy Density Physics and Technology of Ministry of Education, Sichuan University, Chengdu 610065, China}
	
	\author{Jian-Song Pan}
	\email{panjsong@scu.edu.cn}
	\affiliation{College of Physics, Sichuan University, Chengdu 610065, China}
	\affiliation{Key Laboratory of High Energy Density Physics and Technology of Ministry of Education, Sichuan University, Chengdu 610065, China}
	
\begin{abstract}	
		
	While the interaction-induced breakdown of Aharonov-Bohm (AB) cage is typically attributed to uniform bound-pair transport, systems with inhomogeneous exchange interactions realized with Rydberg synthetic lattices exhibit more complex dynamics~\cite{chen2025interaction}. Employing the evolution-path symmetry (EPS) framework developed recently~\cite{pan2026understanding}, we analyze the two-particle dynamics via path interference in Fock space. We find that a homogeneous nearest-neighbor exchange interaction cannot break the AB cage, regardless of whether the long-range exchange interaction is present or not. In contrast, we demonstrate that inhomogeneous nearest-neighbor exchange interaction breaks the destructive-interference EPS, and lifts the degeneracy of many-body compact localized states, thereby generating non-local dispersive eigenstates. Consequently, the initial state gains a non-zero overlap with these dispersive states, enabling delocalized transport. Furthermore, while long-range exchange interaction alone preserves the AB cage, its coupling with nearest-neighbor inhomogeneous exchange interaction opens non-canceling pathways that alter the diffusion profile. Our work connects microscopic path interference with macroscopic spectral reorganization, offering an analytical understanding of the mechanism underlying exchange-interaction-induced transport in Rydberg synthetic lattices.
	
\end{abstract}

\maketitle

\section{Introduction}
	
	The interplay between interactions and artificial gauge fields governs a variety of quantum phases and many-body dynamical behaviors, including the fractional quantum Hall effect~\cite{tsui1982two,laughlin1983anomalous,stormer1999fractional}, topological superfluidity~\cite{wu2013unconventional,zhang2013topological,qu2013topological,wang2017topological}, and interaction-induced quantum transport~\cite{vidal2000interaction,vidal2001disorder,dalibard2011colloquium,tovmasyan2013geometry,celi2014synthetic,greschner2015spontaneous,
	an2017direct,tai2017microscopy,tovmasyan2018preformed,cartwright2018rhombi,di2019nonlinear,danieli2021nonlinear,danieli2021quantum,martinez2023flat,
	zhou2023observation,pan2025interaction,zhao2025observation,chen2025interaction}. In flat-band systems featuring Aharonov-Bohm (AB) cage, this coupling manifests as distinct dynamical behaviors~\cite{vidal1998aharonov,leykam2018artificial}. In the noninteracting limit, localization in clean systems can be achieved through the destructive interference of quantum pathways, facilitated by the joint action of lattice geometry and gauge fields~\cite{sutherland1986localization,abilio1999magnetic,naud2001aharonov}. The fundamental feature of such flat-band systems is that destructive interference flattens the single-particle energy bands, causing the single-particle eigenstates to degenerate into compact localized states (CLSs)~\cite{aoki1996hofstadter,flach2014detangling,longhi2014aharonov,leykam2018artificial,zhang2020compact}. To systematically construct flat-band Hamiltonians, universal flat-band generators have been developed by solving inverse eigenvalue problems with CLSs as a basis~\cite{maimaiti2017compact,rhim2019classification,maimaiti2019universal,maimaiti2021flat,chen2023decoding}. In these flat-band systems, a localized initial wave packet composed of a superposition of CLSs remains strictly confined within a bounded spatial domain and does not diffuse over time~\cite{vidal1998aharonov,li2022aharonov,li2025engineering}. As a classic realization of flat-band lattices, the rhombic lattice with a $\pi$ flux per plaquette exhibits AB cage~\cite{vidal1998aharonov,mukherjee2018experimental}, where multiple tunneling pathways encircling each plaquette undergo destructive interference, thereby suppressing single-particle transport.
	
	The introduction of interparticle interactions fundamentally alters this geometric confinement landscape~\cite{vidal2000interaction,vidal2001disorder}. Conventional theories often attribute such interaction-induced transport to the bound-state transport mechanism~\cite{vidal2000interaction,vidal2001disorder,tovmasyan2013geometry,qin2014statistics,preiss2015strongly,ke2017multiparticle,cartwright2018rhombi,tovmasyan2018preformed,lin2020interaction,kuno2020interaction,pelegri2020interaction,danieli2020many,danieli2021quantum,pan2025interaction}. For instance, in superconducting circuit experiments, a tightly bound pair of particles (doublon), acting as a composite object, experiences an effective gauge flux that is the sum of the individual particle fluxes (for a $\pi$-flux rhombic lattice, the total flux becomes $2\pi$)~\cite{vidal2000interaction,martinez2023flat}. This $2\pi$ flux nullifies the $\pi$ flux responsible for the AB phase, thereby enabling the bound pair to escape the cage.
	
	Recently, in experiments utilizing the highly controllable platform of Rydberg-atom synthetic dimensions~\cite{ozawa2019topological,kanungo2022realizing,chen2024strongly,lu2024wave,trautmann2024realization}, the exchange-interaction-driven breakdown of AB cage and the resulting one-dimensional many-body transport have been observed in flat-band lattices~\cite{chen2025interaction}. Unlike the conventional cases, interparticle coupling in Rydberg synthetic dimensions is dominated by inhomogeneous and non-local dipolar exchange interactions~\cite{browaeys2016experimental,chen2025interaction}. Experiments have demonstrated that as the interaction strength varies, the system exhibits three distinct dynamical regimes: cage preservation, delocalized transport, and a dynamical freeze-out driven by strong interaction-induced disorder~\cite{sierant2017many,chen2025interaction}. Such complex many-body dynamics cannot be captured by the simple uniform bound-state models, which are typically applicable for on-site interactions. The mechanism by which exchange interactions break the AB cage remains unclear. A generic understanding framework insensitive to microscopic details ~\cite{kolley2015strongly,khomeriki2016landau,uchino2016analytical,greschner2016symmetry,di2019nonlinear} is thus highly desirable.
	
	In this work, we systematically investigate the two-particle dynamics in a synthetic rhombic lattice~\cite{chen2025interaction}, analyzing how different components of the dipolar exchange interaction drive the breakdown of AB cage. To elucidate the symmetric mechanism of the interaction-driven breakdown of AB cage, we employ the evolution-path symmetry (EPS) framework developed recently~\cite{pan2026understanding}. While in the noninteracting limit spatial symmetry organizes quantum pathways into pairs with opposite phase factors, preserving the destructive-interference EPS and maintaining AB cage~\cite{vidal1998aharonov,pan2026understanding}, the introduction of interparticle exchange interactions disrupts this interference landscape. By microscopically dissecting low-order evolution-path trees in Fock space, we show that an inhomogeneous nearest-neighbor (NN) exchange interaction disrupts the amplitude balance between mirror evolution pathways due to the spatial variation of exchange coupling strengths, directly breaking the evolution-path symmetry. This symmetry breaking generates non-local dispersive eigenstates, allowing the initial state to gain non-zero overlap with these extended states and initiating delocalized transport. Although the inhomogeneous long-range (LR) exchange alone preserves the AB cage, its coupling with the NN inhomogeneity opens up additional non-canceling long-range hopping pathways that modulate the overall quantum interference. This cooperative mechanism explains the discrepancies in the diffusion profiles of the wave functions between the full dipolar model and the pure inhomogeneous NN model.
	
	The EPS framework maps many-body dynamics onto the coherent superposition of evolution pathways in the temporal domain of the multi-particle Fock space by performing an infinitesimal slicing of the time propagator. For time-independent Hamiltonians, we find that this destructive path interference in the time domain can be physically mapped to a spectral analysis of the system eigenstates. This spectral mapping reveals that the localization properties of the system are fundamentally determined by the projection of the initial state onto the eigenstates: when the initial state projects exclusively onto many-body CLSs localized within a specific spatial domain $\Omega$, the time-evolving wave packet remains strictly localized within $\Omega$ at all times, suppressing delocalized transport~\cite{vidal1998aharonov,vidal2000interaction,pan2026understanding}. Our work elucidates the symmetric mechanism of interaction-driven transport in Rydberg atomic platforms, establishes a clear connection between microscopic path interference and macroscopic spectral reorganization, and provides a useful analytical framework for understanding and controlling quantum transport in many-body spaces via evolution-path symmetry.
	
	The remainder of the paper is organized as follows. In Sec.~II, we introduce the Rydberg synthetic rhombic lattice model and the breakdown of AB cage by presenting the numerical results for the two-particle dynamics under different exchange interaction components. We also present the numerical simulation results of the transport for different types of interactions. In Sec.~III, we introduce the EPS framework for the many-body system and analyze the Fock-space path trees to identify the critical order at which the path-level destructive interference is broken by the inhomogeneous NN exchange interaction. We extend the EPS analysis to the full dipolar model and demonstrate how the coupling between the LR exchange and the NN inhomogeneity alters the diffusion profile. In Sec.~IV, we establish the connection between the microscopic path-interference picture and the spectral projection onto CLSs, showing that the breakdown of AB cage corresponds to the penetration of the initial state into non-local dispersive eigenstates. Finally, we summarize our findings in Sec.~V.
	
\section{Rydberg synthetic rhombic lattice and interaction-induced breakdown of the Aharonov-Bohm cage}
	
	\begin{figure*}[tbp]
		\centering
		\includegraphics[width=0.99\textwidth]{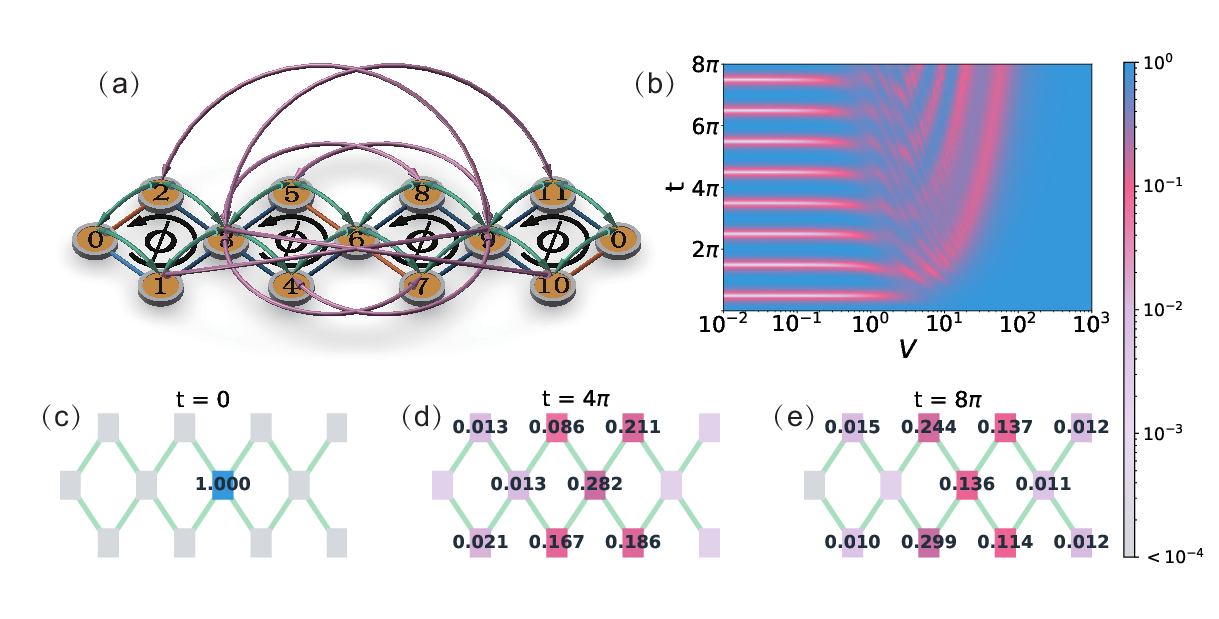}
		\caption{
			(a) Schematic of the model. Solid lines denote single-particle transitions with phases $0$ (dark blue) and $\pm \phi$ (vermilion). Double-headed arrows represent nearest-neighbor (turquoise) and long-range (magenta) exchange interactions.
			(b) Time evolution of the initial-site population $P_6$ versus interaction strength $V$ (for initial state $|6\rangle_A|6\rangle_B$, $J=1/2$, and $\phi = \pi$), illustrating three distinct dynamical regimes: cage preservation ($V \lesssim 10^0$), delocalized transport ($10^0 \lesssim V \lesssim 10^2$), and dynamical freeze-out ($V \gtrsim 10^2$).
			(c)–(e) Spatial population distribution of particle $A$ at $V=10$. Light green lines indicate the lattice geometry, and site labels represent particle populations. Note: panels (b)–(e) share the same color scale.
		}
		\label{fig:model}
	\end{figure*}
	We consider a synthetic dimension lattice consisting of 12 Rydberg states, indexed by $l \in \{0, 1, \dots, 11\}$~\cite{kanungo2022realizing,chen2024strongly,chen2025interaction}. The single-particle tight-binding Hamiltonian $\hat{H}_0$ can be expressed as:
	
	\begin{align}
		\hat{H}_0 = & J \Big(
		(\hat{a}_1^\dagger \hat{a}_0 + e^{i\phi} \hat{a}_0^\dagger \hat{a}_2 + \hat{a}_3^\dagger \hat{a}_1 + \hat{a}_2^\dagger \hat{a}_3) \nonumber\\
		& + (\hat{a}_4^\dagger \hat{a}_3 + \hat{a}_3^\dagger \hat{a}_5 + \hat{a}_6^\dagger \hat{a}_4 + e^{i\phi} \hat{a}_5^\dagger \hat{a}_6) \nonumber\\
		& + (e^{i\phi}\hat{a}_7^\dagger \hat{a}_6 + \hat{a}_6^\dagger \hat{a}_8 + \hat{a}_9^\dagger \hat{a}_7 + \hat{a}_8^\dagger \hat{a}_9) \nonumber\\
		& + (\hat{a}_{10}^\dagger \hat{a}_9 + \hat{a}_9^\dagger \hat{a}_{11} + e^{i\phi} \hat{a}_0^\dagger \hat{a}_{10} + \hat{a}_{11}^\dagger \hat{a}_0)
		\Big) + \text{H.c.}
	\end{align}
	where $J$ represents the transition strength, and $\hat{a}_l$ is the annihilation operator. As shown in Fig.~\ref{fig:model}(a), the dark blue and vermilion solid lines represent transition terms with phases of $0$ and $\pm \phi$, respectively.
	
	{In the experiment~\cite{chen2025interaction}, two distinguishable particles ($A$ and $B$) are considered. Then the total Hamiltonian is given by
		\begin{equation}\label{eq:exchange interaction}
			\hat{H} = \hat{H}_0^A + \hat{H}_0^B + \hat{H}_{\text{int}},
		\end{equation}
		where $\hat{H}_{\text{int}}=\sum_{i>j} V_{i,j} \hat{\chi}_{i,j}$ with $\hat{\chi}_{i,j} = \hat{a}_j^{A\dagger} \hat{a}_i^{B\dagger} \hat{a}_i^A \hat{a}_j^B$ represents the exchange interaction. Here $V_{i,j} = V_{j,i} = \frac{C_3^{i,j}}{C_3^{6,7}} V$ represents the exchange interaction strength between the site pair $(i,j)$. The spatial inhomogeneity is characterized by the $C_3$ coefficients, which depend on the configuration of the synthetic dimension lattice in real space (i.e., it depends on energy levels involved)~\cite{browaeys2016experimental,chen2025interaction}.
	}
	
	Since both the system Hamiltonian and the initial state $|6\rangle_A|6\rangle_B$ possess exchange symmetry with respect to particles $A$ and $B$, the dynamical behaviors of the two particles are identical. Therefore, in the subsequent discussion and spatio-temporal representations, we present the dynamics of particle $A$ as a representative. Fig.~\ref{fig:model}(b) displays the evolution of the initial site population $P_6$ as a function of the interaction strength $V$. When $V \lesssim 10^0$, the AB cage is preserved, and the particles remain localized. In the range $10^0 \lesssim V \lesssim 10^2$, the cage is broken, and the particles undergo transport, as depicted in Fig.~\ref{fig:model}(c)–(e). For $V \gtrsim 10^2$, the system exhibits a freeze-out of $P_6$ dynamics, and the particles regain their localization~\cite{chen2025interaction,sierant2017many,danieli2020many}.
	
	\begin{figure*}[tbp]
		\centering
		\includegraphics[width=0.99\textwidth]{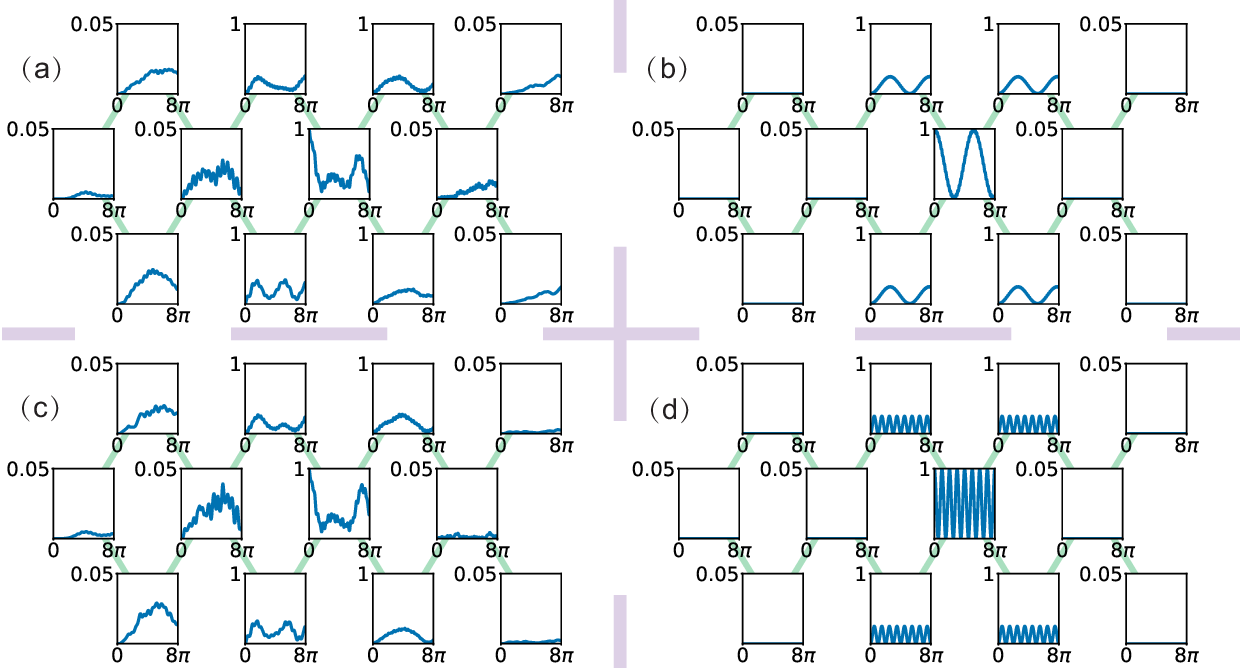}
		\caption{Time evolution of the population $P_i$ of particle $A$ at each site for the initial state $|6\rangle_A|6\rangle_B$, with parameters $J=1/2$, $\phi = \pi$, and $V=10$. The light green lines represent the lattice structure, and each subfigure shows the evolution of the population $P_i$ at the corresponding site over time $t$. The models include: (a) dipolar exchange interaction, (b) uniform NN exchange interaction, (c) inhomogeneous NN exchange interaction, and (d) inhomogeneous LR exchange interaction. The AB cage is broken in (a) and (c), whereas it is preserved in (b) and (d).}
		\label{fig:evolution}
	\end{figure*}
	
	To analyze the mechanism by which the exchange interaction breaks the AB cage, we numerically solve the time-dependent Schrödinger equation (TDSE) to compare the effects of different interaction components on the dynamical evolution. { We decompose the total interaction term $\hat{H}_{\text{int}}$ into nearest-neighbor (NN) and long-range (LR) components:
		
		\begin{equation}
			\hat{H}_{\text{int}} = \hat{H}_{\text{int}}^{\text{NN}} + \hat{H}_{\text{int}}^{\text{LR}}
		\end{equation}
		The nearest-neighbor and long-range exchange interaction terms (indicated by the turquoise and magenta double-headed arrows in Fig.~\ref{fig:model}(a)) are respectively expressed as:
		\begin{align}
			\hat{H}_{\text{int}}^{\text{NN}} & =  \sum_{i,j \in \text{NN}} V_{i,j} \hat{\chi}_{i,j} \\
			\hat{H}_{\text{int}}^{\text{LR}} & =  \sum_{i,j \in \text{LR}} V_{i,j} \hat{\chi}_{i,j}.
		\end{align}
		In general, $V_{i,j}$ depends on the indices $i$ and $j$. For the homogeneous exchange interactions $\hat{H}_{\text{int},Hom}$, $\hat{H}_{\text{int},Hom}^{\text{NN}}$ and $\hat{H}_{\text{int},Hom}^{\text{RL}}$discussed below (for comparison), we take $V_{i,j}$ to be a constant.
	}
	
	Fig.~\ref{fig:evolution} illustrates the density evolution of the particles under various interaction models. Under the full dipolar exchange interaction $\hat{H}_{\text{int}}$, particles diffuse beyond the initial unit cell, indicating a breakdown of the AB cage, as shown in Fig.~\ref{fig:evolution}(a). To identify the dominant components driving this transport, we examine simplified models. Under a uniform NN exchange interaction $\hat{H}_{\text{int},Hom}^{\text{NN}}$, the AB cage is preserved, and no transport occurs [Fig.~\ref{fig:evolution}(b)]. However, when NN inhomogeneity is introduced, the AB cage is broken, and the particles undergo diffusion [Fig.~\ref{fig:evolution}(c)]. This demonstrates that the inhomogeneity of the NN exchange interaction is a necessary condition for breaking the AB cage.
	
	Under an inhomogeneous LR exchange interaction $\hat{H}_{\text{int}}^{\text{LR}}$, the AB cage remains intact, and the particles remain localized, as shown in Fig.~\ref{fig:evolution}(d). Comparing Fig.~\ref{fig:evolution}(a) with Fig.~\ref{fig:evolution}(c), the wave function evolutions do not coincide, indicating that the LR term couples with the inhomogeneous NN term to yield a different diffusion profile. While the LR interaction alone is insufficient to break the local cage [Fig.~\ref{fig:evolution}(d)], its coupling with the inhomogeneous NN term—once the latter initiates particle leakage—enables the participation of long-range pathways, thereby altering the spatial distribution of the diffusing wave packet.
	
	This transport driven by inhomogeneous interactions is distinct from the conventional picture where an interaction-induced bound state experiences a doubled effective magnetic flux. In the following section, we introduce EPS framework to analyze the underlying physical mechanism.
	
\section{EPS framework and discussion}
	
	To gain a deeper understanding of the interaction-driven transport phenomena observed in this study, we employ the recently proposed EPS framework \cite{pan2026understanding} to explain the physical mechanism of localization breakdown from the perspective of path interference. Let us first review this method.
	
	First, consider the single-particle tight-binding Hamiltonian:
	
	\begin{equation}
		\hat{h} = \sum_{i,j} J_{i,j} \ket{j} \bra{i}
	\end{equation}
	where, for $i = j$, $J_{i,i}$ represents the local energy of the particle at site $i$, and for $i \neq j$, it denotes the coupling strength between different sites.
	
	In this representation, the physical process describing a particle evolving from an initial site $i_0$ to a target site $j_0$ is given by the propagator:
	\begin{equation}
		K(j_0, t; i_0, t_0) = \bra{j_0}  U(t, t_0) \ket{i_0}
	\end{equation}
	The propagator $K$ determines the probability amplitude of finding the particle at position $j_0$ at time $t$. Considering a time-independent Hamiltonian, the evolution operator is expressed as $ U(t,t_0) = \exp\left[ -\frac{i}{\hbar} \hat{h} (t - t_0) \right] $. To quantitatively describe the evolution process, we divide the time interval $t-t_0$ into $N$ slices, each with a duration of $\Delta t = (t-t_0)/N$. In the limit $N \to \infty$ (i.e., $\Delta t \to 0$), the single-step evolution operator can be approximated via Taylor expansion:
	
	\begin{equation}
		u(\Delta t) \approx I - \frac{i\Delta t}{\hbar} \hat{h}
	\end{equation}
	
	By multiplying the $N$ individual evolution steps, the total evolution operator can be expanded into a series:
	\begin{align}
		U(t, t_0) &=  [u(\Delta t)]^N \nonumber\\
		&\approx  \sum_{n=0}^N C_N^n \cdot I^{N-n} \cdot \left( -i\frac{\Delta t}{\hbar} \right)^n  \hat{h}^n
	\end{align}
	
	Consequently, the propagator $K(j_0, t; i_0, t_0)$ can also be expanded in a series form as $K(j_0, t; i_0, t_0) = \sum_{n=0}^N K^{(n)}(j_0, t; i_0, t_0)$, where:
	
	\begin{align}
		K^{(n)}(j_0, t; i_0, t_0) &= C_N^n \left( -i\frac{\Delta t}{\hbar}  \right)^n \bra{j_0} \hat{h}^n \ket{i_0} \nonumber \\
		&= C_N^n \left( -i\frac{\Delta t}{\hbar}  \right)^n \xi^{(n)}(j_0,i_0)
	\end{align}
	
	In the continuous-time limit $N \to \infty$, the product of the binomial coefficient and the time step power converges to the standard Taylor expansion coefficient, $\lim_{N\to\infty} C_N^n (\Delta t)^n = \frac{(t-t_0)^n}{n!}$, rendering the discrete path picture mathematically self-consistent with the continuous TDSE.
	
	The physical process of transitions at different orders can be intuitively understood by introducing the propagation factor $\xi^{(n)}(j_0,i_0) = \bra{j_0} \hat{h}^n \ket{i_0}$. For $n=0$, we have $\xi^{(0)} = \bra{j_0} \hat{h}^0 \ket{i_0} = \braket{j_0|i_0} = \delta_{j_0, i_0}$, which corresponds to a zero-step transition and equals 1 only when the starting and ending sites coincide ($j_0 = i_0$). For $n=1$, $\xi^{(1)} = \bra{j_0} \hat{h} \ket{i_0} = J_{i_0, j_0}$ describes the transition amplitude for the particle to directly reach the target site $j_0$ from the initial site $i_0$ via a single step, which represents the direct coupling strength between the two points. For a general $n$-order process, $\xi^{(n)}(j_0,i_0) = \bra{j_0} \hat{h}^n \ket{i_0}$ describes the total propagation amplitude for the particle to reach the target site $j_0$ from the initial site $i_0$ after exactly $n$ steps. Physically, this is equal to the sum of the amplitudes of all possible paths with exactly $n$ steps, where the amplitude of a single path is given by the product of $n$ successive transition amplitudes, exhibiting the product structure $\underbrace{J_{i_0, p} \cdots J_{q, j_0}}_{n\text{ coupling factors}}$. The final factor $\xi^{(n)}(j_0,i_0)$ is the coherent superposition of all these path amplitudes. Notably, if the shortest distance between sites $i_0$ and $j_0$ is $n_{\text{min}}$, then $\xi^{(n)} = 0$ for any step number $n < n_{\text{min}}$, meaning that the lowest-order term contributing to the propagator $K(j_0,t;i_0,t_0)$ starts at order $n_{\text{min}}$.
	
	In the single-particle case, the particle cannot traverse the $\pi$-flux rhombic lattice because the propagator at all orders undergoes complete destructive interference (i.e., $\xi^{(n)} = 0$ for all $n$), leading to a vanishing total propagator $K = 0$~\cite{pan2026understanding}. Taking the shortest path as an example, there are two shortest paths to cross the rhombic lattice, passing through the upper and lower branches, respectively. Under the influence of the $\pi$ flux, the amplitudes of these two paths have opposite signs, resulting in a vanishing lowest-order contribution $\xi^{(n_{\text{min}})} = 0$. Based on the symmetry of the system, higher-order transition paths involving more complex loops ($n > n_{\text{min}}$) also feature a pairwise cancellation of opposite phases, ensuring that $K^{(n)} \equiv 0$ for any $n$. This coherent cancellation of paths at all orders underlies the localization mechanism.
	
	We further generalize the above discussion to the multi-particle case. Since the mathematical form of the many-body Hamiltonian in the Fock-space basis is identical to that of the single-particle Hamiltonian, $\hat{H} = \sum_{\mathbf{i},\mathbf{j}} J_{\mathbf{i},\mathbf{j}} \ket{\mathbf{j}} \bra{\mathbf{i}}$, the series expansion of the propagator can be expressed in the same form:
	
	\begin{align}
		K^{(n)}(\mathbf{j}_0, t; \mathbf{i}_0, t_0) &= C_N^n \left( -i\frac{\Delta t}{\hbar}  \right)^n \bra{\mathbf{j}_0} \hat{H}^n \ket{\mathbf{i}_0} \nonumber \\
		&= C_N^n \left( -i\frac{\Delta t}{\hbar}  \right)^n \Xi^{(n)}(\mathbf{j}_0,\mathbf{i}_0)
	\end{align}
	where $\mathbf{i}_0 = (i_0, m_0, \dots)$ and $\mathbf{j}_0 = (j_0, n_0, \dots)$ represent the initial and target occupational configurations, respectively. In this representation, the many-body propagation factor $\Xi^{(n)}(\mathbf{j}_0,\mathbf{i}_0)$ describes the coherent superposition of all $n$-step evolution paths connecting the initial configuration $\mathbf{i}_0$ to the target configuration $\mathbf{j}_0$ on the many-body Fock-space lattice.
	
	In the two-particle system of this study, to intuitively demonstrate how the exchange interaction induces transport through the breakdown of interference, we systematically investigate the evolutionary dynamics connecting the initial state $\mathbf{i}_0 = (6,6)$ to the target state $\mathbf{j}_0 = (9,9)$ based on the EPS framework.
	
\begin{figure}[tbp]
		\centering
		\includegraphics[width=0.49\textwidth]{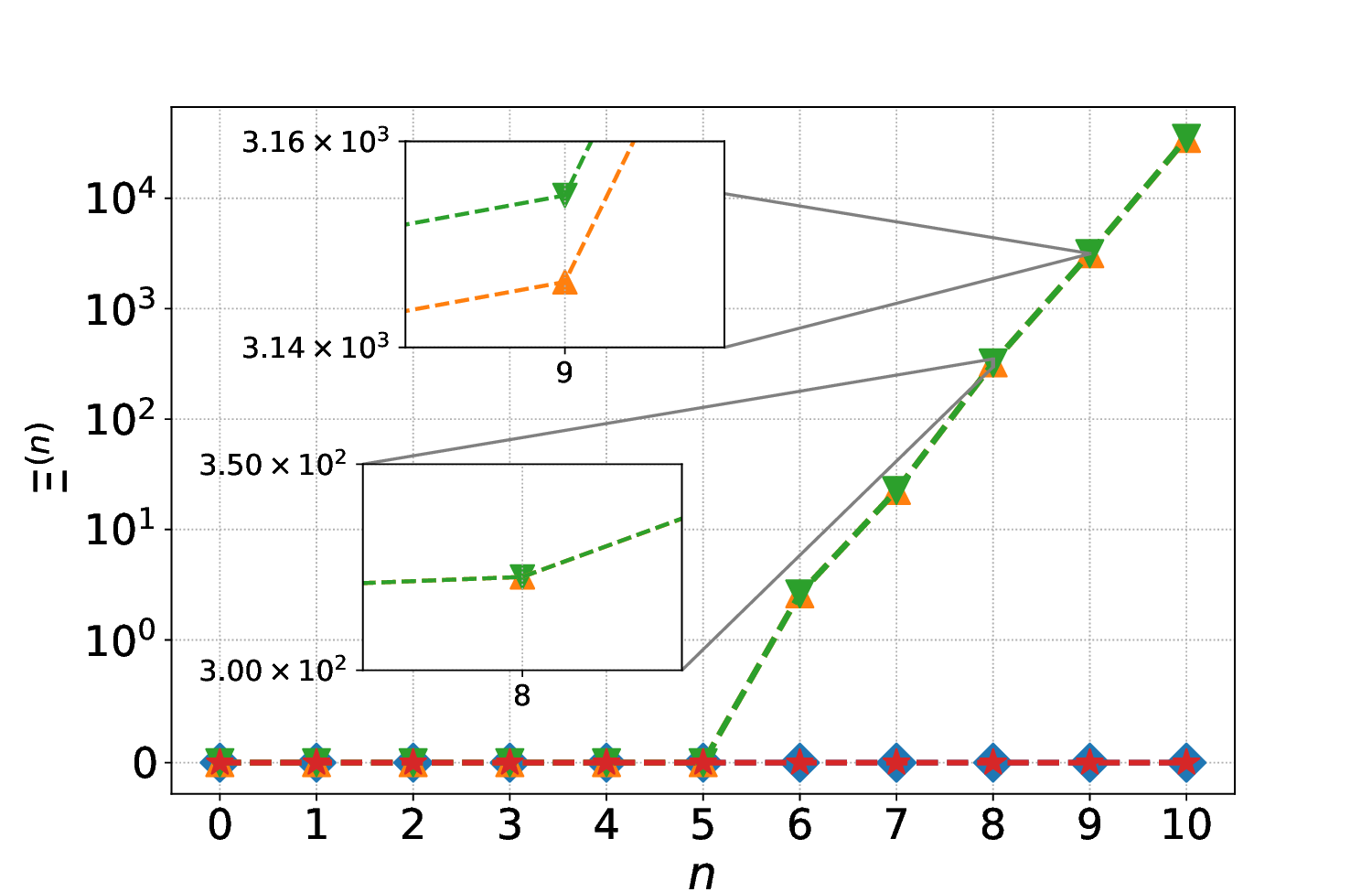}
		\caption{Numerical calculation of the propagation factor $\Xi^{(n)}$: the inhomogeneous LR exchange (blue diamonds) and the uniform NN exchange (red stars) remain at zero within the observation range. In contrast, the inhomogeneous NN exchange (orange upward triangles) and the dipolar exchange (green downward triangles) exhibit an abrupt jump at $n \ge 6$, and they bifurcate after $n=9$ due to the manifestation of the LR contributions.}
		\label{fig:propagator}
	\end{figure}

	The evolution of the numerically calculated propagation factor $\Xi^{(n)}$ as a function of the order $n$ is shown in Fig.~\ref{fig:propagator} ($V = 10$). The results show that for both uniform NN exchange and inhomogeneous LR exchange, $\Xi^{(n)}$ remains zero within numerical error. In the uniform NN case, since the phase corrections are symmetric with respect to the spatial mirror paths, the EPS is preserved at each order. Consequently, the mirror paths cancel out due to the $\pi$ phase difference, and localization is maintained. For the independent inhomogeneous LR term, since it does not act directly on the core links of the cage (site 6 and its nearest neighbors), low-order symmetric paths undergo complete pairwise cancellation, leading strictly to $\Xi^{(n)} = 0$. Although very high-order terms are not explicitly calculated, even if non-canceling contributions were to emerge at high orders, their actual dynamical impact on the propagator is heavily suppressed by the factorial factor $\frac{(t-t_0)^n}{n!}$, thereby keeping the AB cage robustly preserved within the observation time. Thus, low-order paths still tend to cancel in pairs, and its destructive effect on localization only manifests at extremely high orders, leaving the AB cage stable within the observation time. In contrast, $\Xi^{(n)}$ for both the inhomogeneous NN exchange and the dipolar exchange exhibits a sharp increase starting from $n=6$. As the order increases, due to the participation of the LR term, the dipolar exchange and the inhomogeneous NN exchange bifurcate at order $n=9$, which corresponds to the difference in the diffusion characteristics between the two models. This is consistent with the observations in Fig.~\ref{fig:evolution}.
	
\begin{figure*}[tbp]
		\centering
		\includegraphics[width=0.99\textwidth]{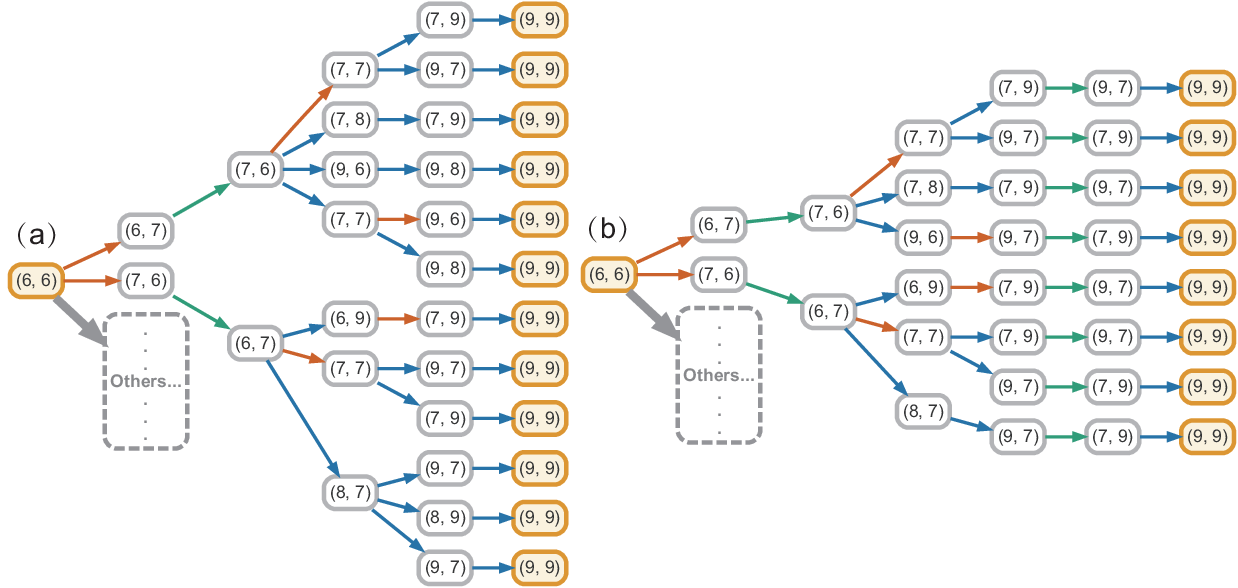}
		\caption{Fock-space path tree in the presence of NN exchange interactions. Nodes $(S_A, S_B)$ represent the locations of the two particles. Arrow definitions: dark blue/vermilion represent single-particle transitions with phase $0/\pi$, respectively; turquoise represents NN exchange transitions. (a) For $n=5$, paths involving $J_{6,7}$ still undergo destructive interference. (b) For $n=6$, paths involving the cooperative effect of $J_{6,7} \& J_{7,9}$ break the destructive interference, leading to the breakdown of localization.}
		\label{fig:tree_near}
	\end{figure*}

\begin{figure*}[tbp]
		\centering
		\includegraphics[width=0.77\textwidth]{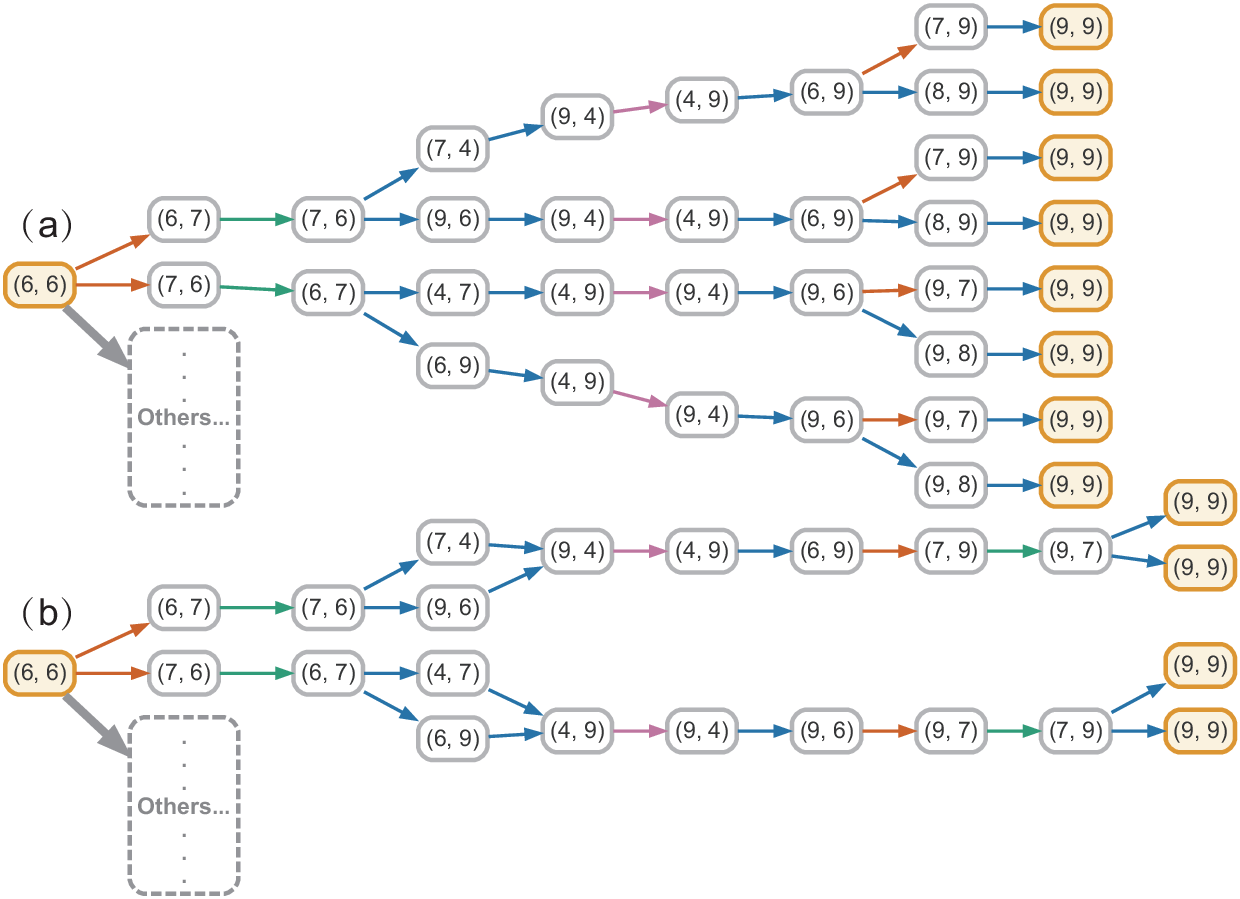}
		\caption{Fock-space path tree in the presence of dipolar exchange interactions. Magenta arrows represent LR exchange transitions, while other arrows are defined identically to those in Fig.~\ref{fig:tree_near}. (a) For $n=8$, the additional paths involving LR terms still undergo destructive interference, and the behavior of the dipolar model is identical to that of the NN model. (b) For $n=9$, the LR paths no longer cancel out, leading to discrepancies between the diffusive wave functions of the dipolar and the NN models.}
		\label{fig:tree_dipole}
	\end{figure*}

	To reveal the underlying microscopic mechanism, we analyze the Fock-space path trees [Figs.~\ref{fig:tree_near} and \ref{fig:tree_dipole}]. The arrow colors in the figures are defined as follows: vermilion/dark blue arrows represent single-particle transitions with phases of $\pi/0$ (providing $-J/J$ coupling terms); turquoise arrows represent NN exchange terms with coupling amplitude $J_{\alpha,\beta}$ (to maintain consistency with the theoretical derivation above, we use $J_{\alpha,\beta}$ to replace the exchange interaction strength $V_{\alpha,\beta}$ of the Hamiltonian in our path-tree analysis); and magenta arrows represent LR exchange terms, which are similarly assigned corresponding coupling amplitudes $J_{\alpha,\beta}$. Since $J_{\alpha,\beta}$ is inhomogeneous, a sufficient condition for global destructive interference is that all subsets of paths containing specific exchange operations cancel out independently. Based on this, we can investigate the interference characteristics of typical path subsets to reveal the microscopic mechanism behind the breakdown of the overall evolution balance:
	
	Fig.~\ref{fig:tree_near}(a) illustrates all paths involving the $J_{6,7}$ operation at order $n=5$ (the minimum order at which the NN exchange participates in the evolution). The calculation shows that this subset still satisfies the destructive interference condition:
	$$\Xi^{(5)}_{\supset J_{6,7}} = (+1+1-1-1+1-1+1+1+1-1-1-1)J^4 J_{6,7} = 0.$$
	When the order increases to $n=6$, the path subset involving the cooperative effect of $J_{6,7}$ and $J_{7,9}$ [Fig.~\ref{fig:tree_near}(b)] no longer cancels out:
	$$\Xi^{(6)}_{\supset J_{6,7} \& J_{7,9}} = (+1+1-1+1+1+1+1-1)J^4 J_{6,7} J_{7,9} \neq 0.$$
	This demonstrates that the inhomogeneous NN exchange interaction begins to destroy the path-level destructive interference starting at the 6th order.
	
	For the dipolar model containing LR terms, its additional contribution relative to the NN-only model originates from paths involving LR exchange. As shown in Fig.~\ref{fig:tree_dipole}(a), at $n=8$, the path subset involving $J_{4,9}$ still maintains destructive interference:
	$$\Xi^{(8)}_{\supset J_{6,7} \text{\&} J_{4,9}} = (+1-1+1-1+1-1+1-1)J^6 J_{6,7} J_{4,9} = 0.$$
	This implies that at and before the 8th order, the LR terms do not introduce differences between the dipolar and the NN-only models. However, when the order reaches $n=9$ [Fig.~\ref{fig:tree_dipole}(b)], the LR term couples with the NN terms, causing this subset to no longer cancel:
	$$\Xi^{(9)}_{\supset J_{6,7} \text{\&} J_{7,9}\text{\&} J_{4,9}} = (+1+1+1+1)J^6 J_{6,7} J_{7,9} J_{4,9} \neq 0.$$
	
	A cooperative mechanism underlies this behavior. In the absence of NN inhomogeneity, the particles are strictly confined to the initial cage, preventing them from accessing the distant site pairs coupled by the LR exchange; thus, the LR terms remain dormant and cage is preserved. However, once the NN inhomogeneity breaks the local interference at $n=6$, the resulting wave-packet leakage populates intermediate sites, enabling the participation of the LR exchange channels. Consequently, the particles can traverse alternative pathways involving long-range exchange (such as the $J_{4,9}$ coupling). At $n=9$, the constructive addition of these LR-mediated pathways ($+1+1+1+1$ in the path subset) modifies the overall interference of the many-body states. Ultimately, such microscopic synergy accounts for the discrepancy in the wave function evolution between the full dipolar and the pure inhomogeneous NN models.
	
	Consequently, the sudden jump in the propagation factor $\Xi^{(n)}$ pinpoints the critical order at which the many-body path symmetry is broken.
	
\section{Bridging EPS and spectral analysis}
	
	While the EPS framework and Fock-space path trees provide a symmetry-based picture of how cage is broken, a purely pathway-oriented analysis becomes mathematically intractable as the order $n$ increases. To establish a general mathematical framework for the localization-to-transport transition, we must bridge this microscopic path-interference picture with the global eigenspectrum of the system~\cite{di2019nonlinear,danieli2021nonlinear,pan2026understanding}.
	
	Fundamentally, the localized cancellation of quantum pathways is deeply connected to the localization properties of the system's eigenstates. In the uniform interaction limit, the destructive path interference across all orders manifests spectrally as many-body CLSs~\cite{tovmasyan2018preformed,danieli2020many,danieli2021quantum}. The introduction of spatial inhomogeneity in the exchange interactions breaks this path-level symmetry, which spectrally translates into the emergence of non-local dispersive eigenstates. Consequently, the breakdown of path-level cancellation is fundamentally equivalent to the penetration of the initial state into these non-local states.
	
	To analyze this spectral reorganization, we solve the Time-Independent Schrödinger Equation (TISE) to obtain the eigenvalues $E_k$ and the corresponding eigenstates $\ket{\varphi_k}$ of the time-independent Hamiltonian $\hat{H}$. The time evolution of the initial state $\ket{\psi(t_0)}$ can then be expressed as:
	\begin{align}
		\ket{\psi(t)} &= U(t,t_0) \ket{\psi(t_0)} \nonumber \\
		&=  U(t,t_0) \sum_k \braket{\varphi_k | \psi(t_0)}  \ket{\varphi_k}	
	\end{align}
	
	For the non-degenerate case, if a contributing eigenstate $\ket{\varphi_k}$ is a many-body CLS strictly localized within a specific spatial region $\Omega$, its time evolution simply acquires a phase factor, $U(t, t_0) \ket{\varphi_k} = e^{-\frac{i E_k (t- t_0)}{\hbar}} \ket{\varphi_k} \approx \sum_{n=0}^N C_N^n \left( -i\frac{\Delta t}{\hbar} \right)^n \hat{H}^n \ket{\varphi_k}$. Here, since $\hat{H}^n \ket{\varphi_k} = E_k^n \ket{\varphi_k}$, the spatial confinement within $\Omega$ is preserved across all expansion orders. Consequently, if all contributing eigenstates $\ket{\varphi_k}$ are many-body CLSs confined within $\Omega$, the time-evolving state $\ket{\psi(t)}$ constructed from their linear superposition will remain strictly localized within $\Omega$ at all times.
	
	For the degenerate case, suppose the energy level $E_d$ corresponds to a set of degenerate eigenstates $ \left\lbrace \ket{\varphi_{d,1}} , \ket{\varphi_{d,2}} , \dots , \ket{\varphi_{d,g}} \right\rbrace $. Even if individual eigenstates generated by exact diagonalization (ED) appear non-local due to arbitrary basis rotations, as long as their linear superposition within the degenerate subspace, $\ket{\varphi_d} = \sum_{\gamma=1}^g \braket{\varphi_{d,\gamma} | \psi(t_0)} \ket{\varphi_{d,\gamma}}$, forms a many-body CLS strictly confined to $\Omega$, its evolution similarly satisfies: $U(t, t_0) \ket{\varphi_d} = e^{-\frac{i E_d (t- t_0)}{\hbar}} \ket{\varphi_d} \approx \sum_{n=0}^N C_N^n \left( -i\frac{\Delta t}{\hbar} \right)^n  \hat{H}^n \ket{\varphi_d} $. Thus, such degenerate components likewise maintain spatial localization without inducing diffusion outside $\Omega$.
	
	Consequently, if the initial state $\ket{\psi(t_0)}$ can be projected onto the following two classes of eigen-subspaces localized within a determined region $\Omega$, the time-dependent state $\ket{\psi(t)}$ will remain strictly localized within $\Omega$ at all times:
	
	(i) non-degenerate many-body CLSs confined to $\Omega$,
	
	(ii) many-body CLSs constructed from linear superpositions of degenerate eigenstates supported on $\Omega$.
	
	To mathematically unify this spectral transformation with the microscopic path-interference picture, we expand the $n$-order many-body propagator $K^{(n)}$ in terms of the eigenstates:
	
	\begin{align}
		\label{eq:eigen_expansion}
		K^{(n)}(\mathbf{j}_0, t; \mathbf{i}_0, t_0) &= C_N^n \left( -i\frac{\Delta t}{\hbar}  \right)^n \sum_k \bra{\mathbf{j}_0} \hat{H}^n \ket{\varphi_k} \braket{\varphi_k | \mathbf{i}_0} \nonumber \\
		&= C_N^n \left( -i\frac{\Delta t}{\hbar}  \right)^n \sum_k E_k^n \braket{\mathbf{j}_0 | \varphi_k} \braket{\varphi_k | \mathbf{i}_0}
	\end{align}
	
	Equation~(\ref{eq:eigen_expansion}) establishes a rigorous link between the spectral (energy) and temporal (path) domains. In a perfectly caged system, the complete destructive path interference across all orders ($K^{(n)}(\mathbf{j}_0, t; \mathbf{i}_0, t_0) \equiv 0$ for any target site $\mathbf{j}_0 \notin \Omega$) is ensured when the initial state $\ket{\mathbf{i}_0}$ projects exclusively onto many-body CLSs (either non-degenerate or degenerate-superposition-formed) that are compactly localized within a bounded spatial domain $\Omega$ containing $\ket{\mathbf{i}_0}$, yielding zero spectral overlap $\braket{\mathbf{j}_0 | \varphi_k} = 0$ at any target site $\mathbf{j}_0$ outside this domain. However, when spatial inhomogeneity in the exchange interactions generates non-local dispersive eigenstates, the initial state gains non-zero projections onto these extended wave functions. Particularly, this corresponds to the emergence of non-zero contributions in the frequency (energy) domain, as indicated by $\braket{\mathbf{j}_0 | \varphi_k} \braket{\varphi_k | \mathbf{i}_0} \neq 0$. When transformed back to the temporal domain, these non-zero frequency-domain terms act as uncompensated (uncancelled) path contributions that disrupt the destructive quantum interference of the path trees, resulting in $K^{(n)} \neq 0$ and the consequent breakdown of the EPS.
	
\begin{figure}[tbp]
		\centering
		\includegraphics[width=0.49\textwidth]{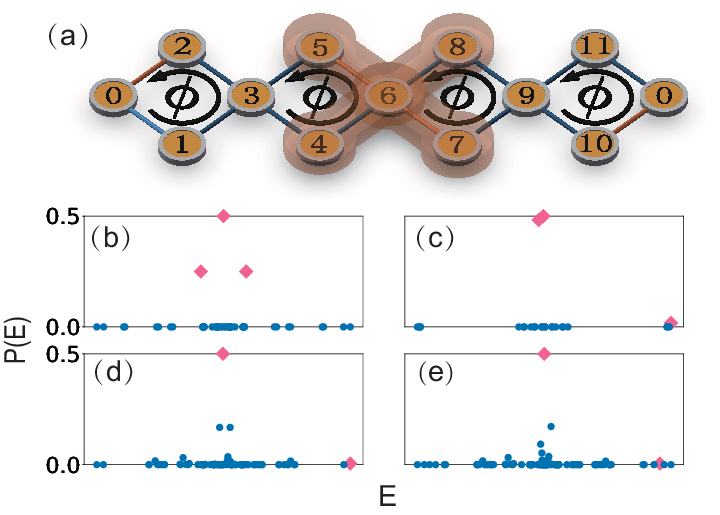}
		\caption{(a) Schematic of the caged region $\Omega = \{4, 5, 6, 7, 8\}$ for a two-particle state. (b)–(e) Projection $P(E)$ of the initial state $|6\rangle_A|6\rangle_B$ at each eigenenergy $E$ under (b) inhomogeneous LR, (c) uniform NN, (d) inhomogeneous NN, and (e) dipolar exchange interactions, respectively, with interaction strength $V = 10$. The vermilion diamonds represent energy levels whose projected components form two-particle CLSs (either non-degenerate CLSs or degenerate-superposition-formed CLSs satisfying Eq.~(\ref{eq:CLS_expression})), whereas dark blue circles represent non-local dispersive states. The results demonstrate that the projections in (b) and (c) are concentrated exclusively on two-particle CLSs, preserving the AB cage, whereas the projections in (d) and (e) scatter into non-local states, breaking the AB cage.}
		\label{fig:projection}
	\end{figure}

	We now focus on the two-particle sector. Fig.~\ref{fig:projection}(a) schematically illustrates the spatial domain of a two-particle state in our lattice. Mathematically, any two-particle state satisfying the compact spatial support condition within the caged region $\Omega = \{4, 5, 6, 7, 8\}$ can be expressed in the general form:
	\begin{equation}
		\ket{\Psi_{\text{CLS}}^{\Omega}} = \sum_{i, j \in \Omega} C_{i, j} \ket{i}_A \ket{j}_B
		\label{eq:CLS_expression}
	\end{equation}
	where $C_{i, j}$ are the expansion coefficients satisfying the normalization condition $\sum_{i, j \in \Omega} |C_{i, j}|^2 = 1$. Crucially, Eq.~(\ref{eq:CLS_expression}) defines the spatial support condition for two-particle states confined to $\Omega$. Such a state constitutes a genuine two-particle CLS if and only if it corresponds to either a non-degenerate eigenstate obtained from ED or a linear superposition of degenerate ED eigenstates compactly supported on $\Omega$.
	
	Figs.~\ref{fig:projection}(b)–(e) show the projection $P(E)$ of the initial state $|6\rangle_A|6\rangle_B$ onto the eigenstates under different interaction models. For the inhomogeneous LR exchange [Fig.~\ref{fig:projection}(b)] and the uniform NN exchange [Fig.~\ref{fig:projection}(c)], the initial state projects exclusively onto two-particle CLSs (either non-degenerate CLSs or degenerate-superposition-formed CLSs) confined within the caged region $\{4, 5, 6, 7, 8\}$ [as defined by Eq.~(\ref{eq:CLS_expression})], which ensures that the wave function remains strictly localized within this bounded domain and the AB cage is preserved. In contrast, for the inhomogeneous NN exchange [Fig.~\ref{fig:projection}(d)] and the dipolar exchange [Fig.~\ref{fig:projection}(e)], the initial state projections scatter into non-local dispersive states lying outside this localized subspace, causing the absence of path-level cancellation and the breakdown of the AB cage. This explains the discrepancies in the transport dynamics observed in Fig.~\ref{fig:evolution}.
	
\section{Conclusion}
	
	In summary, we have systematically investigated the microscopic physical mechanism behind the exchange-interaction-driven breakdown of the AB cage in a Rydberg synthetic rhombic lattice. By combining two-particle TDSE simulations with the EPS framework, we dissect how individual components of the state-dependent dipolar exchange interaction govern the many-body transport dynamics.
	
	Our Fock-space path-tree analysis reveals a clear order-by-order picture of quantum interference. The spatial inhomogeneity of NN exchange interactions serves as the essential trigger for caging breakdown, disrupting the amplitude balance between mirror evolution pathways starting at the 6th evolution order ($n=6$). In contrast, LR exchange alone is insufficient to break caging because its couplings remain dormant during early evolution when particles are localized. However, once NN inhomogeneity initiates wavepacket leakage, LR exchange participates at the 9th order ($n=9$), opening additional non-canceling pathways that distinctly alter the spatial diffusion profile.
	
	Furthermore, we establish a formal theoretical connection mapping the temporal EPS path interference onto the spectral eigenspectrum of the system. We demonstrate that the preservation of path-level cancellation mathematically guarantees that the initial state projects exclusively onto many-body CLSs compactly supported within the caged domain. Conversely, the path-symmetry breaking caused by interaction inhomogeneity manifests spectrally as the emergence of non-local dispersive eigenstates, allowing the initial state to gain overlap with these extended modes and drive delocalized transport.
	
	By unifying microscopic temporal path interference with macroscopic spectral reorganization, our work provides a comprehensive rationale for interaction-driven dynamics that goes beyond conventional uniform bound-pair models. These findings clarify the distinct roles of inhomogeneous NN and LR interactions in Rydberg synthetic dimensions, offering a useful analytical perspective for understanding quantum transport in interacting flat-band systems.
	
\begin{acknowledgments}
	
	This work was supported by the National Natural Science Foundation of China (NSFC) under Grant No. 12574297, the Natural Science Foundation of Sichuan Province under Grant No. 2025ZNSFSC0058, the Fundamental Research Funds for the Central Universities under Grant No. YJ202212, the National Key R$\&$D Program of China under Grant No. 2024YFF0508503, and Beijing National Laboratory for Condensed Matter Physics under Grant No. 2025BNLCMPKF025.
	
\end{acknowledgments}

\bibliography{ref}
	
\end{document}